\documentclass[preprint,amsmath,amssymb,aps,prd,amsart]{revtex4-1}
\usepackage{graphicx}
\usepackage{dcolumn}
\usepackage{bm}
\usepackage{url}
\usepackage{hyperref}
\usepackage{listings}
\usepackage{xcolor}

\definecolor{codegreen}{rgb}{0,0.6,0}
\definecolor{codegray}{rgb}{0.5,0.5,0.5}
\definecolor{codepurple}{rgb}{0.58,0,0.82}
\definecolor{backcolour}{rgb}{0.95,0.95,0.92}

\definecolor{dkgreen}{rgb}{0,0.6,0}
\definecolor{gray}{rgb}{0.5,0.5,0.5}
\definecolor{mauve}{rgb}{0.58,0,0.82}
\definecolor{light-gray}{gray}{0.25}

\def\bluecolorifnotalreadymauve{%
    \extractcolorspec{.}\currentcolor
    \extractcolorspec{mauve}\stringcolor
    \ifx\currentcolor\stringcolor\else
        \color{blue}%
    \fi
}

\lstdefinestyle{MMA}{%
    language=Mathematica,%
    breaklines=true,
     backgroundcolor=\color{backcolour},
    basicstyle={\sffamily\footnotesize},
    numbers=left,
    numberstyle=\tiny\color{gray},
    numbersep=5pt,
    breaklines=true,
    captionpos={t},
    frame={lines},    
    rulecolor=\color{black},
    framerule=0.5pt,
    columns=flexible,
    tabsize=2,
    morekeywords={SortBy,RandomReal,Spectrogram,BandpassFilter}
}

\lstdefinestyle{PYTH}{
    language=Python,
    frame={lines},
    backgroundcolor=\color{backcolour},   
    commentstyle=\color{codegreen},
    keywordstyle=\color{magenta},
    numberstyle=\tiny\color{codegray},
    stringstyle=\color{codepurple},
    basicstyle=\ttfamily\footnotesize,
    breakatwhitespace=false,         
    breaklines=true,                 
    captionpos=b,                    
    keepspaces=true,                 
    numbers=left,                    
    numbersep=5pt,                  
    showspaces=false,                
    showstringspaces=false,
    showtabs=false,                  
    tabsize=2
}

\lstdefinestyle{RUBY}{
    language=Ruby,
    frame={lines},
    backgroundcolor=\color{backcolour},   
    commentstyle=\color{codegreen},
    keywordstyle=\color{magenta},
    numberstyle=\tiny\color{codegray},
    stringstyle=\color{codepurple},
    basicstyle=\ttfamily\footnotesize,
    breakatwhitespace=false,         
    breaklines=true,                 
    captionpos=b,                    
    keepspaces=true,                 
    numbers=left,                    
    numbersep=5pt,                  
    showspaces=false,                
    showstringspaces=false,
    showtabs=false,                  
    tabsize=2
}

\begin{document}

\title{Notes on Quantum Soundscapes and Music}

\author{Miles Blencowe}
\email{miles.p.blencowe@dartmouth.edu}
\affiliation{Department of Physics and Astronomy, Dartmouth College, Hanover, New Hampshire 03755, USA}

\author{Michael Casey}
\email{michael.a.casey@dartmouth.edu}
\affiliation{Departments of Music and Computer Science,Dartmouth College, Hanover, New Hampshire 03755, USA}

\author{Kimberly Tan}
\thanks{Present Address: Ministry of Education, Singapore 138675}
\email{ktme4031@gmail.com}
\affiliation{Thayer School of Engineering and Department of Philosophy, Dartmouth College, Hanover, New Hampshire 03755, USA}

\date{\today}

\begin{abstract}

We describe our investigations concerning the sonification of measured data from experiments involving various mesoscopic mechanical oscillator systems cooled down close to their quantum ground states, and music generation from a programmed quantum computer that subjects a single quantum bit (``qubit") to various unitary rotations, composed in order to test for the breakdown of macroscopic realism as expressed by the violation of the Leggett-Garg inequality. ``Listening'' to data via their resulting sonifications facilitates the discovery of signals that might otherwise be hard to detect in common graphic (i.e., visual) representations, and for the quantum computer music experiment provides a complementary way to discern when the measured qubit data violates macroscopic realism with some prior listening training. The resulting soundscapes and music also provide a complementary window into the quantum realm that is accessible to non-experts with open ears.
\end{abstract}

\maketitle

\section{\label{sec:Introduction}Introduction}
The investigations described in this paper began with an exchange between two of the authors (K. T. and M. B.) back in early 2019 on the differences and similarities between musical and scientific instruments, prompted by Ref. \cite{tresch13}. Of specific interest to us are scientific instruments that measure the states of quantum systems, for example mesoscopic mechanical resonator systems cooled close to their quantum ground states (modeled as quantum harmonic oscillators) \cite{guo2019,zhou2019,cattiaux2021} and two-state systems (i.e., qubits \cite{bienfait2019}). The data produced from such experiments is typically in the form of a tab-separated or comma-separated file of measured output power values at different preset frequencies (i.e., a power spectrum), or output power values at different time increments. Taking the inverse Fourier transform of the former type of data set (with random assigned phases) yields a temporal data set that is similar to the latter form. Instead of visually plotting the data to show power versus time as is commonly done, we may instead sample the data at the rate of a few thousands per second, converting it into audio output using any one of several available software packages (such as Mathematica) on a laptop connected to speakers (or headphones). The change in power value from one time instant to the next determines the sound intensity and frequency range, depending on the preselected sampling rate. 

The only ``dial" involved in the above-outlined sonification procedure is the sampling rate, affecting the pitch of a vibrating mesoscopic oscillator, for example. The resulting sound for an oscillator system will not be  a purely sinusoidal tone;  material elastic anharmonicity, defects within the vibrating structure, attachment to a supporting structure and substrate at a finite temperature (in the millikelvin range, for example), noise added by the various amplification stages up to room temperature of the electromagnetic signal transducing the mechanical system motion, as well as the back reaction noise of the amplifier onto the mechanical oscillator, will all contribute to the ``timbre" of a more complex sound. We term such sounds that are produced as a result of the quantum-limited signal transduction and amplification of systems operating in the quantum regime--for example, mechanical oscillators undergoing predominantly quantum zero-point motion while cooled close to their quantum ground state, quantum jumps due to phonon emission/absorption etc.--``quantum soundscapes". This extends to the meso-microscopic quantum domain the usage of the term ``soundscape" that has been extensively applied to describe the everyday sounds we experience while immersed in our classical, macroscopic environment \cite{schafer1994}. While the produced sounds are by necessity classical in order to be audible, they ultimately result from a quantum measurement process; by listening, we gain information about the evolving state of the quantum system. 

The utility of sonifying quantum data has been appreciated before, going back to the Geiger counter dating from early last century, which detects gamma ray photons and other ionizing particles, transducing the electrical detection signals into audio ``clicks"; one such notable demonstration was used to broadcast the ``voice of the atom" from a radio station at the University of Kansas on the evening of May 20, 1926 \cite{cady1926}. Another notable example from earlier this century was the detection of Josephson oscillations in superfluid Helium-4 by listening \cite{hoskinson2005}; the oscillations stand out from a background hissing noise as a tone that starts at the high frequency end of the listening range and falls to the low frequency end in the span of a few seconds.  These two examples point to complementary ways in which the sonification of experimental data from quantum systems can be beneficial: first and foremost, signals may be discovered by listening that would otherwise be hard to notice in the data when represented in the usual visual way; in particular, the human auditory system is quite effective at picking out tonal sounds from a noisy background. Second, sonification provides a largely untapped sensory means through which to expose the quantum realm to the interested, listening non-expert. In a similar spirit to the former, earlier example given above, data from experiments involving superconducting qubits was sonified and performed live to a public audience in New Haven on June 14, 2019 \cite{topel2022}.

In Sec. \ref{sec:qsoundscape}, we describe our independent  quantum soundscape projects \cite{roaf2020,blencowe2020,blencowe2022}. These involved first reaching out to various experimentalists to share their data. In the recording \cite{roaf2020}, sonifications of data from six different experiments involving quantum dynamical processes are presented: (a) a $2.6~{\mathrm{mm}}$ long silicon nanostring with  fundamental vibrational frequency $950~{\mathrm{kHz}}$ undergoing Brownian motion that is cooled in stages down to an average occupation number of a few phonons using light pressure \cite{guo2019}; (b) a $50~\mu{\mathrm{m}}$ long silicon nitride nanowire with  fundamental vibrational frequency $4~{\mathrm{MHz}}$ undergoing Brownian motion at a temperature of a few tens of mK \cite{zhou2019} (a longer recording is given in \cite{blencowe2020}); (c) a superconducting qubit emitting and recapturing surface acoustic phonons \cite{bienfait2019}; (d) superfluid helium thermal phonons getting converted into photons that are detected individually \cite{patil2022}; (e) photons detected individually from a diode laser; (f) X-ray photons registered by a balloon detector launched in Antarctica, with the photons produced by relativistic electrons streaming from the Sun \cite{millan2021}. In recording \cite{blencowe2022}, a $10~\mu{\mathrm{m}}$ diameter aluminium disc vibrating at around $15~{\mathrm{MHz}}$ was cooled to a faction of a mK in a nuclear demagnetisation cryostat, such that the disc was in its vibrational quantum ground state \cite{cattiaux2021}. The quantum soundscape recordings \cite{roaf2020,blencowe2020,blencowe2022} contain tonal and percussive elements, along with ``noise" characterised by a broad spectral power density. Recordings (c), (d), and (e) of \cite{roaf2020} are exclusively percussive in nature, reflecting the detection of discrete energy quanta (i.e., photons or phonons). In contrast, recordings (a) and (b) of \cite{roaf2020} and recording \cite{blencowe2022} are largely tonal, reflecting the continuous vibrational nature of the mechanical oscillators undergoing quantum Brownian motion.

Two separate signals were discovered by listening to these soundscapes; the signals had not been noticed previously when visualizing the experimental data. In recording (e) of \cite{roaf2020}, a rapid transient tonal burst in the x-ray photon detection rate was heard in what was otherwise a noisy signal with slowly varying amplitude. In recording \cite{blencowe2022}, a sudden increase in the disc oscillator frequency was heard at one point, before more slowly returning to the original frequency tone. The former signal may have been caused by the high energy tail of a pulsating aurora \cite{millan2020}, while the latter signal was possibly due to a small temperature increase fluctuation of the disc oscillator, the cause of which is unclear \cite{collin2022}. Both described signals are hard to spot in the original, graphically displayed data, without first knowing where to look; however, the signals are readily apparent in the sonification of the data.

While the transduced, amplified, and sonified electrical signals in the just-described recordings arise from microscopic, evolving quantum systems, there are no distinctly identifiable quantum signatures in the resulting sounds; the number distribution of detected photon clicks in a given time window for recordings (d) and (e) in \cite{roaf2020} is well-fit by a Poisson distribution \cite{blencowe2020b,patil2022}, as are numerous other macroscopic, classical discrete random sonic processes, such as the sound of hail hitting a metal roof. And recording \cite{blencowe2022} of a mechanical oscillator cooled down to close to its quantum vibrational ground state would not distinctly differ (beyond the lower amplitude of the oscillator tone)  from the sound of the same oscillator undergoing  classical, thermal Brownian motion at a much higher temperature. 

In light of this sonic similarity between the recorded quantum and everyday classical random sounds, a natural question arises as to whether a more ``quantum" sound can be generated? More concretely: can we ``play"  a scientific instrument (that is controlling
and measuring a microscopic quantum system) like a
musical instrument, generating a form of distinctly quantum ``music"?
In Sec. \ref{sec:qmusic}, we describe just such a quantum music project, which involves programming a physical quantum computer to execute a particular quantum process. The sequence of measured output `$\pm 1$'s are converted into selected note combinations using a digital music synthesizer; by listening to the resulting ``composed music", the ear can be trained to pick out quantum versus classical correlations, such as to be able to distinguish quantum from classical processes in separately generated data sets that have the same temporal correlation functions as the original, training data set.

Our composed music \cite{blencowe2024} provides an aural test of the Leggett-Garg inequality \cite{leggett1985,emary2014}, which quantifies, through a  
combination of measured two-time correlation functions involving a system with a two-valued variable, whether the system obeys a macroscopic realist (i.e., classical) versus quantum description; as a sort of temporal Bell inequality, the Leggett-Garg inequality is in our view well-suited to form the quantum mechanical basis of quantum music. The system that we consider is an evolving qubit described by the Hamiltonian $H=\frac{1}{2}\hbar\Omega \sigma_x$ (with $\hbar$ the reduced Planck constant, $\Omega$ a frequency parameter that characterizes how rapidly the qubit evolves, and $\sigma_x$ the Pauli-X operator), which is repeatedly prepared in the initial $\sigma_z$ (Pauli-Z operator) eigenstate with eigenvalue $-1$ and with $\sigma_z$ measured after a preset time interval \cite{javadi2024}. This procedure is experimentally implemented using an IBM quantum computer that is programmed via Qiskit \cite{javadi2024}. Depending on whether the measured qubit state is $+1$ or remains $-1$, an ascending or descending musical scale is generated via the digital music synthesizer Sonic Pi \cite{aaron}. 

The generated music \cite{blencowe2024} 
contains random elements arising from the intrinsically probabilistic nature of quantum mechanics--it has neither an identifiable beginning nor an ending; if the piece is played too slowly, then our mind's ear is unable to fill in the musical notes to come, no matter how long the piece has been playing; the notes are not perfectly correlated temporally. On the other hand, if the piece is played too quickly, then we lose the tonality of the piece. However, if the piece is played in some intermediate tempo range,  can the musical mind's ear process the imperfect correlations due to the randomness by listening for a sufficiently long duration, and hence without an expert knowledge of quantum mechanics (or music) distinguish the latter from macroscopic realism by hearing the violation of the Leggett-Garg inequality? This is the precise question that motivates our quantum music investigation.

The use of (mathematical) physics principles to generate music with random elements (often via programmable computers) has been explored previously, most notably by Iannis Xenakis in what he termed  ``Free Stochastic Music" \cite{xenakis}, dating from the 1950s.
Various ideas concerning ``Quantum Music" have been explored much more recently, in part motivated by the advent of readily accessible basic quantum computers; see, for example, Refs. \cite{sturm2001,weimer2010,quantummusic,putz2017,medic2018,vedral2018,kirke2018,garner2018,molmer2018,helweg2018,medic2018b,marletto2018,yamada2021,miranda2022,miranda2024,abdyssagin2024,loncar2024}. Our approach is closest to that of Ref. \cite{yamada2021}, which seeks to distinguish intrinsic quantum random process from classical deterministic processes for the example of a two-level atom undergoing Rabi oscillations and spontaneous decay; certain tones and harmonics are assigned, depending on which level the atom is in. Defining what is ``quantum music"  in our view is very much an open question at present, with considerable space (and time) for exploration; just as for classical (in the physics sense) music, we anticipate that quantum music will be limitless and diverse in its possibilities. 

In Sec. \ref{sec:qsoundscape}, we describe how the soundscape \cite{blencowe2022} was produced as a representative example, while in Sec. \ref{sec:qmusic} we describe how the quantum music piece \cite{blencowe2024} was generated. Sec. \ref{sec:conclude} gives concluding remarks. The appendices contain the codes used to produce the recordings \cite{blencowe2022} and \cite{blencowe2024}

\section{\label{sec:qsoundscape}Quantum Soundscapes}
In this section, we focus on the soundscape recording \cite{blencowe2022}, which was generated with data \cite{data} obtained from the experiment described in Ref. \cite{cattiaux2021}; the data sonification procedure was similar to that applied to obtain the other quantum soundscape recordings \cite{roaf2020,blencowe2020}, serving as an informative, representative example.
\begin{figure}[htb]
\begin{center}
\includegraphics[width=5in]{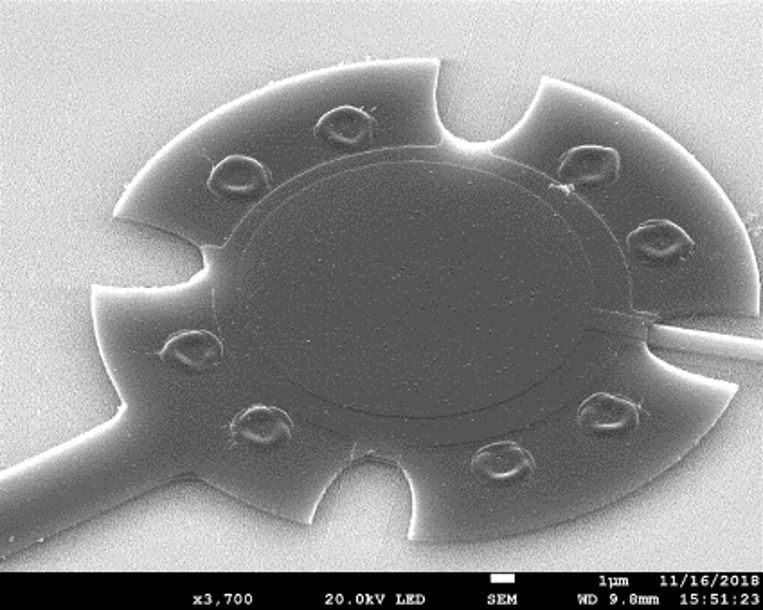} 
\caption{\label{drumfig}Scanning electron microscope image of the drum-shaped, aluminium mechanical oscillator with diameter $15~\mu{\mathrm{m}}$ and thickness $100~{\mathrm{nm}}$ that was investigated in Ref. \cite{cattiaux2021}. (Figure courtesy of L. Mercier de L\'{e}pinay and M. Sillanp\"{a}\"{a}.)}
\end{center}
\end{figure}
The investigated device is an aluminium disc with a $15~\mu{\mathrm{m}}$ diameter and $100~{\mathrm{nm}}$ thickness (Fig. \ref{drumfig}). The disc is separated by a  $50~{\mathrm{nm}}$ vertical gap from an aluminium electrode on a sapphire substrate surface beneath, allowing the disc to freely vibrate in the direction transverse to the substrate surface, much like a vibrating drumhead; the fundamental (i.e., lowest) transverse mode vibrational frequency of the micrometer-sized drumhead is $f_0\approx 15~{\mathrm{MHz}}$, dictated by the built-in tensile stress, the mass density, and the size-dimensions of the aluminium drumhead. Due to the clamping of the drumhead edge boundary to the surface substrate and also to the presence of defects within the aluminium drumhead material, the oscillating drumhead motion will damp out through coupling via the accompanying oscillating strain field. Left undisturbed by any external drive force, the drumhead described classically will undergo random, thermal Brownian motion, with a dominant peak at the fundamental transverse vibrational frequency in the displacement noise spectrum having a linewidth given by the damping rate. Furthermore, the fundamental vibrational frequency will undergo random shifts due to the thermal fluctuating energies of the strain-coupled defects within the aluminium disc. 

The aluminium disc and electrode beneath the disc are located at the opposite, terminal ends of a coiled wire circuit, such as to form a  microwave inductor-capacitor ``cavity" with frequency $f_c\approx 6~{\mathrm{GHz}}$ that couples to the disc motion via the displacing vacuum separation dependence of the disc-electrode capacitance. With the microwave cavity-mechanical oscillator device situated within a nuclear demagnetization fridge that attains a lowest average temperature $T\approx 0.5~{\mathrm{mK}}$ during the running of the experiment, the thermal average phonon occupation number of the fundamental vibrational mode is $n(T)=\left(\exp\left({\frac{h f_0}{k_B T}}\right)-1\right)^{-1}\approx 0.3$, where $h=6.626\times 10^{-34}~{\mathrm{J}}\cdot{\mathrm{Hz}}$ is Planck's constant and $k_B=1.38\times 10^{-23}~{\mathrm{J}}\cdot{\mathrm{K}}^{-1}$ is the Boltzmann constant. Thus, all of the vibrational modes of the drumhead are close to being in their quantum ground states; because of the Heisenberg uncertainty principle, the drumhead cannot be at absolute rest (in contrast to a classical drumhead), but instead undergoes irreducible quantum zero-point motion. This motion is detected through the dependence of the microwave cavity frequency on the drumhead centre of mass position $x$: $f_c(x)=\frac{1}{2\pi\sqrt{LC(x)}}$, where $C(x)$ is the drumhead disc-electrode capacitance and $L$ the coiled circuit inductance. In particular, for the data set sonification recording \cite{blencowe2022}, the microwave cavity is pumped with a so-called ``blue detuned" tone $f_p=f_c+f_0$ (i.e., pump frequency fixed as the sum of the cavity frequency and fundamental mechanical drumhead frequency determined at the higher, $100~{\mathrm{mK}}$ cryogenic temperature device calibration stage), with the pump power corresponding to approximately $600$ photons in the cavity. The resulting reflected microwave power spectrum consists of the central pump peak at $f_p=f_c+f_0$ and mechanical signal sideband satellite peaks at integer multiples of the fundamental vibrational mode frequency $f_0$ relative to the pump frequency; the dominant sideband peak occurs at $f_p- f_0=f_c$ (i.e., at the microwave cavity frequency).

The capacitively coupled microwave cavity thereby serves as a transducer, converting the mechanical signal of the quantum vibrating drumhead into an electromagnetic signal, which is sequentially amplified at two temperature stages, first at liquid Helium temperature ($4~{\mathrm{K}})$ and subsequently at room temperature ($300~{\mathrm{K}}$) (see Appendix A of Ref. \cite{zhou2019} for the microwave experimental setup). The room temperature amplified, dominant microwave sideband signal is then demodulated from the reference pump frequency and utilized to extract the average phonon occupation number of the fundamental vibrational mode of the mesoscopic drumhead \cite{cattiaux2021}. Note that transduction from mechanical to amplified electromagnetic signals also forms the basis of certain electronic musical instruments, such as the electric guitar \cite{mcdonald2007,horton2009}.  

The resulting experimental data \cite{data} that we used for the sonification consists of $11930$ files, each containing 4095 data points in two-column (i.e., tab separated) format giving the measured demodulated voltage amplitude signal versus frequency spectrum in $6.706~{\mathrm{Hz}}$ increments, covering a $27.45~{\mathrm{kHz}}\, (=4094\times 6.706~{\mathrm{Hz}})$ acquisition bandwidth zero-centred at the mechanical frequency $f_0$. Each data file corresponds on average to around 30 seconds of acquisition time, so that the collective data represents an experiment run time of around four days, during when the  microwave cavity-drumhead device sample within the demagnetization fridge starting at a temperature of approximately $15~{\mathrm{mK}}$,  was cooled down to a sustained, average temperature of $0.5~{\mathrm{mK}}$, before finally warming up to around $65~{\mathrm{mK}}$ at the end stage of the run.     

The URL link to the sonification recording of the data is given in  Ref. \cite{blencowe2022}, and the Mathematica \cite{wolfram2020} code used to generate the recording is given in appendix \ref{app:sonodatacode}. A moving average of ten data files (corresponding to around 300 seconds) is first performed and then the noise floor arising from the microwave amplification circuitry is subtracted off from the mechanical signal spectral peak. Because the spectral data lacks phase information (only the voltage amplitudes are recorded), we multiply to each voltage data point the phase factor $e^{i\phi}$, with $\phi$ selected randomly from the interval $(0,2\pi)$. An inverse Fourier transform is then performed on the complex voltage data set spectra to obtain time domain voltage data. This time domain data is ``played" with a sampling rate of $2015~{\mathrm{Hz}}$, selected such that the drumhead motion signal occurs in the audio frequency range between approximately $460$--$520~{\mathrm{Hz}}$. The drumhead signal is enhanced by using a bandpass filter with lower (upper) cutoff frequency $400\, (550)~{\mathrm{Hz}}$. With each inverse Fourier transformed data file comprising $2\times 4095$ data points,  each data set provides approximately four seconds of sound recording time sampled at $2015~{\mathrm{Hz}}$; the full data set comprising 11930 files therefore gives a sound recording time of $4.8\times 10^4$ seconds, i.e., around 13 hours!  

\begin{figure}[htb]
\begin{center}
\includegraphics[width=6in]{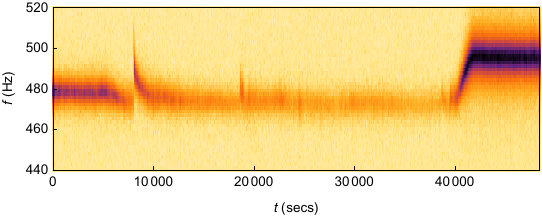} 
\caption{\label{spectrogramfig}Spectrogram of the full data set sonification. The vertical axis gives the sound frequency spectrum in Hertz and the horizontal axis gives the time elapsed in seconds. The colour represents sound amplitude, with deep purple to black denoting ``very loud" and light yellow denoting ``very quiet". }
\end{center}
\end{figure}
It would be challenging to listen to the whole of such a long recording; in order to aid in signal discovery, a spectrogram was made of the full generated recording, shown in Fig. \ref{spectrogramfig}. Noticeable features are the decrease in sound amplitude and frequency of the vibrating drumhead during the initial cooling down stage, and the correspondingly larger increase in sound amplitude and frequency as the vibrating drumhead warms to the higher final temperature towards the end of the recorded experimental run. During the sustained lowest temperature cooling period, when the drumhead is at a temperature of around $0.5~{\mathrm{mK}}$ corresponding to a $0.3$ average phonon occupation number, the resulting, almost quantum zeropoint motion of the drumhead ``rings" at around $470~{\mathrm{Hz}}$. This frequency was chosen through the selected time domain data sampling rate, inspired by the sounds of bees in a field recording made by one of us (M. B.) \cite{blencowe2022b}--a purely aesthetic choice. 

Also clearly noticable is a switching event occurring at around the 8025 second mark, when the drumhead frequency (and sound amplitude) suddenly increases dramatically to around $500~{\mathrm{Hz}}$, and subsequently decreases more slowly. This switching event can be easily heard, once we know when it occurs; recording \cite{blencowe2022} is a very short, 129 second  excerpt of the full sound recording, selected around this switching event. This event was not noticed in the original data set, with its sudden onset occurring within the acquisition time of just a  single voltage versus frequency spectrum data file; it would be hard to detect such an event without knowing of its existence a priori, given that there are 11920 such data files (the proverbial needle in a haystack)! The origin of this and other less prominent frequency switching events seen in the acoustic spectrogram (Fig. \ref{spectrogramfig}) are likely due to temperature fluctuations; as with other mechanical vibrational systems at low temperatures, two-level system (TLS) defects coupling via the mechanical strain to the vibrational mode of interest affect the frequency of the latter as the TLS population temperature changes \cite{cattiaux2021}. The cause(s) of the temperature fluctuations are however unknown \cite{collin2022}.

\section{\label{sec:qmusic}Quantum Music}
In this section, we compose a form of quantum music that is played on a quantum musical instrument. In particular, our musical instrument is a quantum computer which executes a particular program, generating a string of measured output ones and zeros that are converted into musical notes via the use of a pre-programmed music synthesizer. The resulting music is quantum in the sense that it could not have been produced by playing a classical (in the physics sense) musical instrument (other than a simulation--see comment later below). 

We codify ``classicality" here through the two basic assumed principles that macroscopic systems satisfy as quoted from Ref. \cite{leggett1985}:
\begin{enumerate}
   \item[(A1)]{Macroscopic realism (MR): A macroscopic system with two or
more macroscopically distinct states available to it will at all times be in one or the other of these states.}
   \item[(A2)]{Noninvasive measurability (NIM) at the macroscopic
level: It is possible, in principle, to determine the state of the system with arbitrarily small perturbation on its subsequent dynamics.}
\end{enumerate}
Consider now a two-state system with observable $Q$ taking either of the two values $Q_i=\pm 1$ at time $t_i$. Supposing that we have an ensemble of such systems, all prepared in some identical state at time $t_0<t_1<t_2\dots$, then we define the two-time correlation function  as $C_{12}=\sum_{Q_1, Q_2=\pm 1} Q_1 Q_2 P_{12}(Q_1,Q_2)$, where $P_{12}(Q_1,Q_2)$ is the joint probability of obtaining the value $Q_1$ at time $t_1$ and the value $Q_2$ at time $t_2>t_1$; in short, we can think of the two-time correlation function as the ensemble average $\langle Q_1 Q_2\rangle$. However, principle (A1) also allows the two-time correlation function to be obtained from the three-time correlation function--for example: $P_{13}(Q_1,Q_3)=\sum_{Q_2=\pm 1}P_{123}(Q_1,Q_2,Q_3),\, t_1<t_2<t_3$, such that $C_{13}=\sum_{Q_1,Q_2,Q_3=\pm 1}Q_1 Q_3 P_{123}(Q_1,Q_2,Q_3)$. Obtaining the two-time correlation functions $C_{12}$, $C_{23}$, and $C_{13}$ in this way, the following inequality involving these correlation functions can be derived \cite{leggett1985,emary2014}:
\begin{equation}
-3\leq K\leq 1,
\label{keq}
\end{equation}
where $K=C_{12}+C_{23}-C_{13}$. Equation (\ref{keq}) is known as the Leggett-Garg (LG) inequality for a two-state system that is measured at three distinct times \cite{leggett1985}.

\begin{figure}[htb]
\begin{center}
\includegraphics[width=5in]{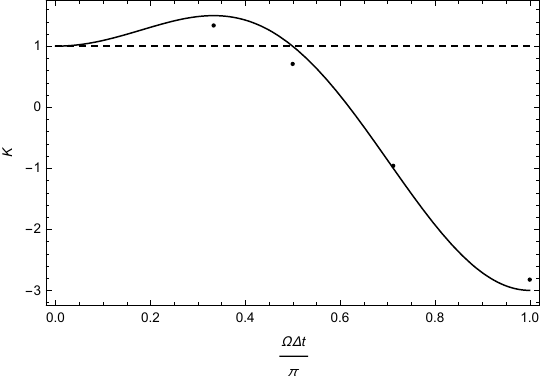} 
\caption{\label{kplotfig}Plot of the theoretical prediction for $K_{\mathrm{theor}}=C_{12}+C_{23}-C_{13}=2 \cos\left(\Omega\Delta t\right)-\cos\left(2\Omega\Delta t\right)$ versus $\Delta t$ (solid black line) for a qubit evolving according to the Hamiltonian $H=\frac{1}{2}\hbar \Omega \sigma_x$ and subject to measurements of the observable $\sigma_z$. The experimental $K_{\mathrm{exp}}$ results (black dots) for an actual qubit (realized through one of IBM's quantum computers), are evaluated at the selected time intervals $\Delta t=\pi/(3\Omega),\, \pi/(2\Omega),\, 0.712\pi/\Omega,\, \pi/\Omega$. Data points above the horizontal dashed line violate the LG inequality (\ref{keq}).}
\end{center}
\end{figure}
A quantum two-state system (i.e., a ``qubit") described by the Hamiltonian $H=\frac{1}{2}\hbar \Omega \sigma_x$ and with measurement observable $Q=\sigma_z$ (where $\hbar$ is Planck's constant, the $\sigma_i$ are the standard Pauli matrices, and $\Omega$ is a parameter which governs the rate at which the qubit state evolves), has the two-time correlation function \cite{emary2014}
\begin{equation}
    C_{i j}=\cos\left[\left(\Omega(t_j-t_i\right)\right], 
    \label{qcorreq}
\end{equation}
independent of the initial state at time $t_0<t_i<t_j$. Taking equal time intervals between measurements: $t_2-t_1=t_3-t_2=\Delta t$, we have $K=2 \cos\left(\Omega\Delta t\right)-\cos\left(2\Omega\Delta t\right)$. For the measurement time interval $\Delta t=\pi/\Omega$, $K$ attains its minimum value $-3$, consistent with the LG inequality (\ref{keq}). On the other hand,  for the measurement interval $\Delta t=\pi/(3\Omega)$, $K$ attains its maximum value $+1.5$, which violates the LG inequality (\ref{keq}); classical, macroscopic realism therefore breaks down for a range of measurement intervals in the case of the qubit evolving under the above Hamiltonian. The solid line curve in Fig. \ref{kplotfig} gives $K$ versus $\Delta t$ following from the theoretical prediction for the qubit system two-time correlation functions (\ref{qcorreq}). The region above the horizontal dashed line corresponds to the time interval range $0<\Delta t<\pi/(2\Omega)$ for which the LG inequality is violated. 

The dots in Fig. \ref{kplotfig} are experimental data that was obtained from one of IBM's quantum computers (``Brisbane") during August 20-21, 2024. Instead of performing measurements over an ensemble of identically prepared qubits, the measurements were repeated on the same qubit, prepared in the same initial state each time with value $Q=-1$. Table \ref{table:k} lists the selected measurement interval times $\Delta t$, the corresponding experimental $K_{\mathrm{exp}}$ values obtained after averaging over $500$ repeated measurements, as well as the  theoretical predicted $K_{\mathrm{theor}}$ values. 
\begin{table}[h!]
\centering
\begin{tabular}{||c c c||} 
 \hline
 $\Omega\Delta t/\pi$ & $K_{\mathrm{exp}}$ & $K_{\mathrm{theor}}$\\ [0.5ex] 
 \hline\hline
 1/3 & 1.339 & 1.5\\ 
 1/2 & 0.709 & 1\\
0.712 & -0.96 & -1\\
1 & -2.824 & -3\\ [1ex] 
 \hline
\end{tabular}
\caption{Selected $\Delta t$ values used in the experiment, along with the resulting $K_{\mathrm{exp}}$ values obtained after averaging over $500$ repeated measurements. Also shown for comparison are the predicted $K_{\mathrm{theor}}$ values.}
\label{table:k}
\end{table}
While not as large as the theoretical prediction, the LG inequality is significally violated experimentally for the time interval $\Delta t=\pi/(3\Omega)$. The Python code  that was used to run the quantum computer via Qiskit \cite{javadi2024}, and generate the data presented in Fig. \ref{kplotfig} and Table \ref{table:k}, is given in appendix \ref{app:qccode}.

In order to turn the quantum computer into a musical instrument, we must convert the measured output data from the former into a sequence of musical notes using a synthesizer; the quantum computer with connected synthesizer may then be said to form a musical instrument which can be played (i.e., programmed) such as to produce quantum music. In particular, the assumed principles (A1) and (A2) quoted above can be recast in musical language:
\begin{enumerate}
   \item[(A1)]{Macroscopic realism: A played macroscopic musical instrument with two or
more macroscopically distinct sounds (i.e., notes) available to it will at all times be producing one or the other of these notes.}
   \item[(A2)]{Noninvasive listening at the macroscopic
level: It is possible, in principle, to hear the note of the musical instrument with arbitrarily small perturbation on its subsequent played notes.}
\end{enumerate}
The two-time correlation functions defined above can now be applied to played note sequences generated by the musical instrument, where the joint probability of one note following another is averaged over many repetitions. The goal is then to discern through listening whether the LG inequality (\ref{keq}) is violated, such that one or both of the above assumed principles (A1) and (A2) in musical form do not hold, signalling that the played music is non-classical in the physics (i.e., macroscopic realism violating) sense. 

While a performing trio playing the keyboards, for example, or a standard laptop computer with a suitable software program can produce LG violating music that is indistinguishible from that generated by the synthesizer-connected quantum computer, the former are simulations (or ``metaphors" \cite{raedt2016}) of the latter physical process involving a genuine evolving quantum two state system.         

The (converted) data output resulting from the quantum computer program given in appendix \ref{app:qccode} is in the form of a sequence like the following: $(-1,+1,-1,+1,-1,-1,-1,+1,\dots)$, where the first, third, fifth etc. entries always have the value $Q=-1$ corresponding to the initial prepared state of the qubit, while the second, fourth, sixth etc. entries may take either measured values $Q=\pm 1$ with frequencies of occurence depending on the selected time interval $t_j-t_i$. In particular, the sequence comprises pairs of numbers of the form $(-1,\pm 1)$  which are the result of repeated qubit state preparations that are followed by measurements of $Q$ after the time interval $t_j-t_i$ (termed ``shots"). Since $P_{ij}(+1,-1)=P_{ij}(+1,+1)=0$, the two-time correlation function definition simplifies to $C_{i j}=P_{ij}(-1,-1)-P_{ij}(-1,+1)$, where $P_{ij}(Q_i,Q_j)$ is the joint probability of obtaining the value $Q_i$ at time $t_i$ and $Q_j$ at time $t_j>t_i$. The task of the composer is then to select a particular distinct sound (or note) for the synthesizer to play, depending on whether the measured qubit value is $+1$ or $-1$. The possibilities in making this choice are limitless, reflecting the creative freedom of the composer. We chose the E flat Dorian scale \cite{dorian} (used for example in the composition ``So What"  \cite{sowhat}), desiring a sound that is neither too western nor too eastern. In order to accentuate the differences between the two possible preparation$\rightarrow$measurement outcome pairs: $-1\rightarrow -1$ or $-1\rightarrow +1$, we furthermore chose to descend the E flat Dorian scale in the former case, and ascend the scale in the latter case. For example, if the data output sequence takes the repeated form: $(-1,+1,-1,+1,-1,+1,-1,+1,\dots)$, then we ascend the scale, while on the other hand, if the sequence takes the repeated form: $(-1,-1,-1,-1,-1,-1,-1,-1,\dots)$, then we descend the scale; typically, a data output sequence will take neither of these limiting, perfectly correlated forms, generating a sequence of ascending and descending notes on the E flat Dorian scale. Finally, for the synthesized notes, we use Shepard tones \cite{shepard1964}, so that each note volume sounds the same; regular harmonic tones would otherwise emphasize the higher pitch notes to the listener's ear. 

\begin{figure}[htb]
\begin{center}
\includegraphics[width=5in]{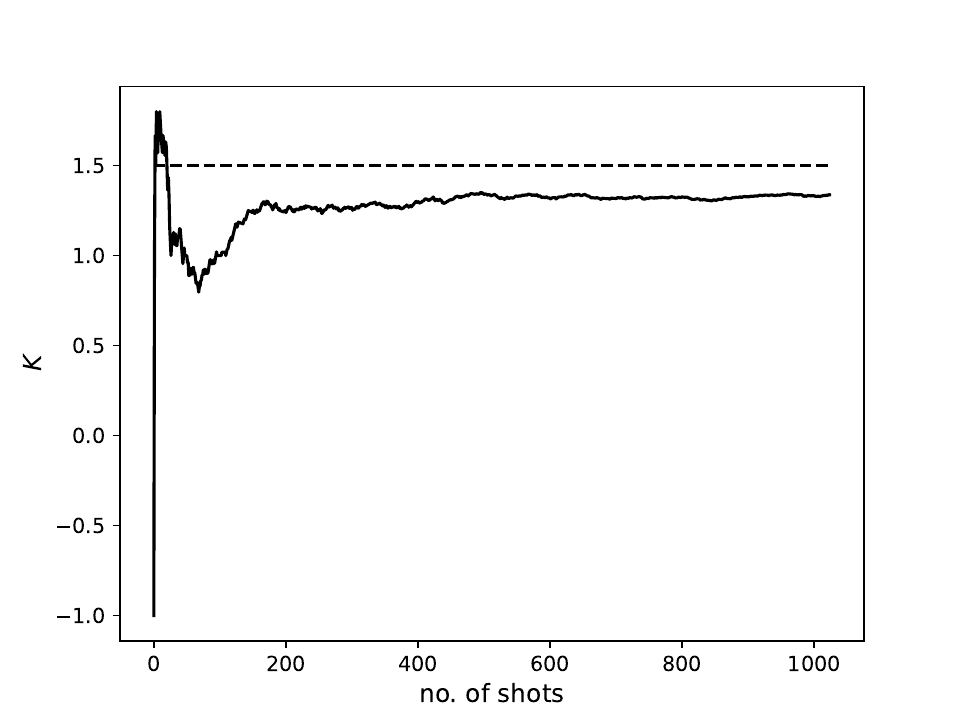} 
\caption{\label{kplot2fig}Plot of the experimental correlation value $K_{\mathrm{exp}}$ for the time interval $\Delta t=\pi/(3\Omega)$ versus the accumulated number of repeated state preparation and measurements (i.e., ``shots") over which the correlation value is calculated. When the number of shots is $500$, we have $K_{\mathrm{exp}}=1.339$. The horizontal dashed line indicates the predicted theoretical value $K_{\mathrm{theor}}=1.5$.}
\end{center}
\end{figure}
By listening to such a note sequence that represents the data set generated for repeated state preparations and measurements separated by the time interval $t_j-t_i$ (i.e., shots), the aim is to gain a musical sense of the two-time correlation function $C_{ij}$.  With the correlation measure $K=C_{12}+C_{23}-C_{13}$ appearing in the LG inequality expression (\ref{keq})  involving two correlation functions corresponding to a measurement interval $\Delta t$ and one corresponding to the measurement interval $2\Delta t$, it is natural to have three  voices making up the composition resulting from three independent data sets, two for the interval $\Delta t$ and one for the interval $2 \Delta t$. We used the same steady tempo for each of the three voices, playing them together as a polyphony. While there is no obvious way to sonically represent the subtraction of $C_{13}$ in the definition of $K$ (since combining voices corresponds most naturally to addition), it may be that a form of subtraction is performed by listening over a period of time. 

In selecting the tempo of the piece, it is informative to examine how the measured correlation $K_{\mathrm{exp}}$ settles on a reasonable steady value as the accumulating number of shots used in the averaging to determine $K_{\mathrm{exp}}$ grows. Figure \ref{kplot2fig} gives the accumulating average $K_{\mathrm{exp}}$ value  for the measurement interval $\Delta t=\pi/(3\Omega)$, where the LG inequality is maximally violated; note that $K_{\mathrm{exp}}$ consistently exceeds one (i.e., violates the LG inequality) beyond around $100$ shots. Thus, hearing the LG violation necessarily requires listening to a musical phrase within the piece comprising over 100 sequential notes, a consequence of the inherent probabilistic randomness of measurement outcomes in quantum mechanics. [This is to be contrasted with a familiar, standard classical (in the physics sense) piece of music, where the notes of the piece are perfectly correlated; often the music can recognized after listening to only a short phrase such as the beginning of the piece.] If the tempo is too slow, then our musical memory forgets the notes that came before as the piece is being played and we cannot hear the LG violating correlations. On the other hand, if the tempo is too fast, then we lose the musicality of the piece. By trial and error, we settled on a tempo of $0.15$ seconds per note, corresponding to requiring around 15 seconds or longer of continuous listening time in order to be able to hear the LG violation.  

The URL link to the quantum music composition recording generated following the procedure described above is given in Ref. \cite{blencowe2024}. The composition consists of four movements, corresponding sequentially to the experimental correlation measure values $K_{\mathrm{exp}}=-2.824$, $-0.96$, $0.709$, and $1.339$. By listening to the complete piece a few times in close succession, we have found that it is possible in a later, subsequent listening to distinguish the three ``classical" [i.e., obeying the LG inequality (\ref{keq})] movements from the ``quantum" [i.e., violating the LG inequality (\ref{keq})] movement. Furthermore, a subsequent listening may be of a quantum statistical ``variation" of the composition, in the sense that the four movements are generated from another run on a quantum computer, producing a different data set with the same selection of four different measurement time intervals $\Delta t$ as above. The listening need not start at the beginning of a particular movement; having random as well as repetitive elements, the movements (and their quantum statistical variations) have no identifiable introduction or conclusion as in a standard classical, deterministic (in the physics sense) piece of music.

\section{\label{sec:conclude}Conclusion}
We have described here our first investigations into sonifying data from quantum experiments, producing quantum soundscapes and generating quantum music. The soundscapes are quantum in the sense that the physical systems and initial measurement transduction stages are operating close to the quantum limit, registering in some experiments quantum zero-point motion and in other experiments single energy quanta (i.e., phonons or photons). The quantum music composition involves programming a qubit system and converting the output data into musical notes via a synthesizer, such that the temporal correlation function involved in the Leggett-Garg inequality can be measured by listening. 

Moving beyond this initial stage of investigation, it would be desirable to carry out a more rigorous quantum music listening test, which might involve a large group of listeners who are not necessarily music or quantum theory experts. The testing procedure could first involve having the group listen to the quantum music recording \cite{blencowe2024} several times. At a later time, a quantum statistical variation of the composition  (resulting from a different quantum computer experiment data set)  is played to each listener, where the order of each of the four movements is randomized. Each listener is then asked to identify the LG inequality violating movement, without communicating with any other listener during the test.

Another exploration might involve a trio of professional musicians playing a transcription of the recording \cite{blencowe2024} on wind instruments or keyboards (or both), for example. The trio  then improvises variations of the different movements, keeping with the same E flat Dorian scale over one or two octaves, say, and at roughly the same tempo as the recording. An interesting question is whether the improvising trio can play a variation such that the extracted temporal correlation measure $K$ is close to that of the recorded quantum music movement; in particular, can the musicians improvise on the fourth, Leggett-Garg violating movement such as to achieve a temporal correlation measure $1<K\lesssim 1.5$?

The quantum composition recording \cite{blencowe2024} has a constant beat rhythm, with each note in the three polyphonic voices lasting for 0.15 seconds. It would also be interesting to try to compose pieces where correlation information is also encoded in a varying rhythm. Furthermore, richer polyphonies with additional voices might be generated by running quantum computing processes involving multiple qubits that test LG inequalities \cite{lambert2016,ku2020,santini2022}.

During Iannis Xenakis' thesis defense on May 18, 1976 at the Sorbonne, the philosopher and member of the Jury, Olivier Revault d'Allones posed the following connected series of questions in response \cite{xenakis1985}: ``How can we hope to interest scholars and scientists and thereby perceive these new mental structures which Xenakis himself alludes to today? Art's use of science benefits the former more than the latter. Is this lack of balance bad? And if yes, how can we overcome this?" By discovering quantum physics signals in experimental data through listening to the generated quantum soundscapes or music, we provide one possible answer to these questions. At the least, listening provides a complementary means with which the expert and non-expert alike can experience the quantum realm.

\acknowledgments{We thank Andrew Armour, Victoria Aschheim, Shadi Ali Ahmad, Richard Beaudoin, Audrey Bienfait, Cecilia Blencowe, William Braasch Jr., Andrew Cleland, Eddy Collin,  Arij Elfaki,  Viola Gatti Roaf,  Simon Gr\"{o}blacher, Jack Harris, John Kulvicki, Hiroko Kumaki, Robyn Millan, Paul Nation, Franco Nori, Dan Rockmore, Alex Schultz, Keith Schwab, Danielle Simon, Spencer Topel, Steve Weinstein, Daniel Westphal, Lucy Yu, Nicole Yunger Halpern, and Yiren Zheng for very helpful conversations and contributions to the quantum soundscapes and music investigations. M. B. would also like to thank the Department of Physics and Astronomy at Tufts University for the opportunity to present this work in a lively colloquium on November 8, 2024, and an especial thank you to Roger and Tess Concepcion for providing a welcoming and inspiring environment at their Toronto home, where the beginning of the paper was written. This work was supported in part by the NSF under Grant No. PHY-2011382, and by Dartmouth's Neukom Scholars Program and Women in Science Project.}

\appendix
\section{Data Sonification Code}
\label{app:sonodatacode}
This appendix section gives the Mathematica \cite{wolfram2020} code for producing the sonification recording \cite{blencowe2022} using the experiment data files \cite{data}. 
\lstnewenvironment{MMA}{\lstset{style=MMA}}{}
\begin{MMA}
(*Input the data files from the folder "11122020 0.4mK D6" renamed as "0.txt", "1.txt", "2.txt", etc *)
files = SortBy[FileNames["/Users/miles/Desktop/Quantum Music/CattiauxNatComm2021/11122020 0.4mK D6/*.txt"], StringLength];
datacsets = Map[Import[#, "Data"] &, files];

(*Obtain moving average*)
ave = 10;
moveave = Table[1/ave Sum[datacsets[[n + i, 2 ;; 4096, 1 ;; 2]], {i, 0, ave - 1}], {n, 1, Length[datacsets[[;; , 1, 1]]] - ave + 1}];
datac = {};
Do[AppendTo[datac, moveave[[n, ;; , 1]] - Mean[moveave[[n, 1500 ;; 1800, 1]]]], {n, Length[moveave]}];

(*Generate time domain data through Inverse Fourier Transform*)
randphase = RandomReal[{0, 2 Pi}, {Length[datac[[;; , 1]]], Length[datac[[1, ;;]]]}];
temp = datac Exp[I randphase];
tempr = Reverse[Conjugate[temp[[;; , 2 ;; Length[datac[[1, ;;]]]]]], 2];
tempp = Join[temp, tempr, 2];
if = {};
Do[AppendTo[if, InverseFourier[tempp[[n]], FourierParameters -> {1, -1}]], {n, Length[datac]}];
ift = Re[Flatten[if]];

(*Sonify the time domain data*)
sound = BandpassFilter[ListPlay[ift, SampleRate -> 2015, PlayRange -> All], {2513, 3456}]
Export["sound.wav", sound]

(*Generate a spectrogram of the sonification*)
Spectrogram[sound, PlotRange -> {440, 520}, FrameLabel -> {"t (secs)", "f (Hz)"} ]
Export["Users/miles/Desktop/Quantum Music/spectrogram.pdf", %]
\end{MMA}

\section{Experimental Qubit LG Inequality Test Code}
\label{app:qccode}
This appendix section gives the Python code used to generate the experiment data for testing the Leggett Garg inequality on an evolving qubit via IBM's Qiskit \cite{javadi2024}.
\lstnewenvironment{PYTH}{\lstset{style=PYTH}}{}
\begin{PYTH}
import numpy as np
# Imports from Qiskit
from qiskit import QuantumCircuit, transpile
# Imports from Qiskit Runtime
from qiskit_ibm_runtime import QiskitRuntimeService
# To run on hardware, select the backend with the fewest number of jobs in the queue
service = QiskitRuntimeService(channel="ibm_quantum", token="<MY_IBM_QUANTUM_TOKEN>")
#service = QiskitRuntimeService(channel="ibm_quantum")
backend = service.least_busy(operational=True, simulator=False)
backend.name
#Build circuits
dt=1/3*np.pi
#dt=1/2*np.pi
#dt=np.pi
#dt=0.712*np.pi
qc21 = QuantumCircuit(1,1)
qc21.rx(dt,0)
qc21.measure(0,0)
qc32 = QuantumCircuit(1,1)
qc32.rx(dt,0)
qc32.measure(0,0)
qc31 = QuantumCircuit(1,1)
qc31.rx(2*dt,0)
qc31.measure(0,0)
#Draw circuit_21
qc21.draw('mpl')
#Draw circuit_32
qc31.draw('mpl')
#Run on an actual quantum computer
shotnum=1024
compiled_circuit21 = transpile(qc21, backend)
backend.run(compiled_circuit21, shots=shotnum,memory=True)
compiled_circuit32 = transpile(qc32, backend)
backend.run(compiled_circuit32, shots=shotnum,memory=True)
compiled_circuit31 = transpile(qc31, backend)
backend.run(compiled_circuit31, shots=shotnum,memory=True)
#Import generated data
import numpy as np
import json
from qiskit.result import Result
shotnum=1024
d21=json.load(open('/Users/miles/Desktop/Quantum Music/August20_24_Actual_pi_over_3/job21/result.json'))
result21 = Result.from_dict(d21)
data21=result21.get_memory()
data_array21=np.array([int(i) for i in data21])
for i in range(0,2*shotnum-1,2):
    data_array21=np.insert(data_array21,i,0)
ones_array=np.array([1]*2*shotnum)
dataconv_array21=2*data_array21-ones_array
d32=json.load(open('/Users/miles/Desktop/Quantum Music/August20_24_Actual_pi_over_3/job32/result.json'))
result32 = Result.from_dict(d32)
data32=result32.get_memory()
data_array32=np.array([int(i) for i in data32])
for i in range(0,2*shotnum-1,2):
    data_array32=np.insert(data_array32,i,0)
ones_array=np.array([1]*2*shotnum)
dataconv_array32=2*data_array32-ones_array
d31=json.load(open('/Users/miles/Desktop/Quantum Music/August20_24_Actual_pi_over_3/job31/result.json'))
result31 = Result.from_dict(d31)
data31=result31.get_memory()
data_array31=np.array([int(i) for i in data31])
for i in range(0,2*shotnum-1,2):
    data_array31=np.insert(data_array31,i,0)
ones_array=np.array([1]*2*shotnum)
dataconv_array31=2*data_array31-ones_array
#Calculate cummulative K correlation measure
dt=1/3*np.pi
#dt=1/2*np.pi
#dt=0.712*np.pi
#dt=np.pi
data21p=[0]
for i in range(0,2*shotnum-1,2):
    data21p.append(dataconv_array21[i]*dataconv_array21[i+1])
else:
    data21p.remove(0)
c21temp=np.cumsum(data21p)
c21=[0]
for i in range(0,int(2*shotnum/2)):
    c21.append(c21temp[i]/(i+1))
else:
    c21.remove(0)
    data32p=[0]
for i in range(0,2*shotnum-1,2):
    data32p.append(dataconv_array32[i]*dataconv_array32[i+1])
else:
    data32p.remove(0)
c32temp=np.cumsum(data32p)
c32=[0]
for i in range(0,int(2*shotnum/2)):
    c32.append(c32temp[i]/(i+1))
else:
    c32.remove(0)
    data31p=[0]
for i in range(0,2*shotnum-1,2):
    data31p.append(dataconv_array31[i]*dataconv_array31[i+1])
else:
    data31p.remove(0)
c31temp=np.cumsum(data31p)
c31=[0]
for i in range(0,int(2*shotnum/2)):
    c31.append(c31temp[i]/(i+1))
else:
    c31.remove(0)
c21=np.array(c21) 
c32=np.array(c32)
c31=np.array(c31)
kdata=c21+c32-c31
ktheor=2*np.cos(dt)-np.cos(2*dt)
#Plot the cummulative K correlation measure and compare with theoretical K value 
import matplotlib.pyplot as plt
data_plot=plt.plot(kdata,color='black')
theory_plot=plt.hlines(y=ktheor, xmin=0, xmax=1024, linewidth=1.5, color='black',linestyle='--')
plt.xlabel('no. of shots', fontsize=12)
plt.ylabel('K', style='italic', fontsize=12)
plt.savefig("/Users/miles/Desktop/Quantum Music/kplot.pdf")
kdata[500]
#Save measured output data in csv format
np.savetxt("/Users/miles/Desktop/Quantum Music/Aug_20_data21_actual_pi_over_3.csv", data_array21, fmt='%s', delimiter=",")
np.savetxt("/Users/miles/Desktop/Quantum Music/Aug_20_data32_actual_pi_over_3.csv", data_array32, fmt='%s', delimiter=",")
np.savetxt("/Users/miles/Desktop/Quantum Music/Aug_20_data31_actual_pi_over_3.csv", data_array31, fmt='%s', delimiter=",")
\end{PYTH}

\section{Music Synthesizer Code}
\label{app:musiccode}
This appendix section gives the Ruby code used to convert the experimental quantum computer data output into musical note sequences via Sonic Pi \cite{aaron}. 
\lstnewenvironment{RUBY}{\lstset{style=RUBY}}{}
\begin{RUBY}
#Q-Music Synthesis with Shepard tones: K=1.339 movement only
require 'csv'
root_dir = '/Users/miles/Desktop/Quantum Music/'
define :read_csv do |filename|
  data = Array.new
  CSV.foreach(filename){|row|
    row_i = row.map{|r| r.to_i}
    data << row_i
  }
  return data
end
shots=500
start=1
#Generate Shepard tones using code written by Benjamin Wand
#https://github.com/benjaminwand/Sonic-Pi-Shepard-tones-function
define :shepard do |pitch, time = 1, attack = 0.5, release=1, volume = 1, synth = :blade, center = 78|
  use_synth synth
  if pitch.class == Array
    pitch.each do |one_pitch|                   # recursion
      in_thread do
        shepard one_pitch, time, attack, release, volume, synth, center
      end
    end
  else
    (1..9).each do |octave|                     # 9 octaves
      if pitch.class == Symbol
        tones = (note_info pitch, octave: octave).midi_note    # make numbers
      end
      if pitch.class == Float or pitch.class == Integer
        tones = pitch % 12 + 12 * (octave + 1)  # midi numbers of tone
      end
      equal_amp = 30.0 / tones ** 1.5 * volume  # all at same-ish volume
      middle =  (tones - center)/10.7           # centers bell curve, 78 -> [-5 .. 5]
      gauss = equal_amp * 2**-( middle**2/10)   # bell curve * equalized volume
      play tones, amp: gauss, attack: attack, release: release, sustain: time, release: attack
    end
  end
  sleep time
end
#pitch
p= [ :Ds, :F, :Fs, :Gs, :As, :C, :Cs].ring #Dorian Eb
t=0.15 #time
a=0.02 #attack
r=0.02 #release
v21=1.0 #volume
v32=1.0 #volume
v31=2.0 #volume
s=:sine #synth
c=68 #center
t2=0.7 #time
a2=0.2 #attack
r2=0.5 #release
sl=2 #sleep
#K=1.339 data
data21 = read_csv(root_dir+'Aug_20_data21_actual_pi_over_3.csv').flatten
data32 = read_csv(root_dir+'Aug_20_data32_actual_pi_over_3.csv').flatten
data31 = read_csv(root_dir+'Aug_20_data31_actual_pi_over_3.csv').flatten
#Generate the three polyphonic voices
in_thread do
  n1=start
  n21=0
  shots.times do
    puts n1
    shepard p[n21], t, a, r, v21, s, c
    if data21[n1]==0
      n21=n21-1
    else
      n21=n21+1
    end
    n1=n1+2
  end
  shepard p[n21], t2, a2, r2, v21, s, c
end
in_thread do
  n2=start
  n32=0
  shots.times do
    puts n2
    shepard p[n32], t, a, r, v32, s, c
    if data32[n2]==0
      n32=n32-1
    else
      n32=n32+1
    end
    n2=n2+2
  end
  shepard p[n32], t2, a2, r2, v21, s, c
end
n3=start
n31=0
shots.times do
  puts n3
  shepard p[n31], t, a, r, v31, s, c
  if data31[n3]==0
    n31=n31-1
  else
    n31=n31+1
  end
  n3=n3+2
end
shepard p[n31], t2, a2, r2, v21, s, c
\end{RUBY} 
\end{document}